# SISTEMA DE CONTROL BASADO EN EL ESTÁNDAR GRAFCET PARA LA ALTERNACIÓN DE BOMBAS CENTRIFUGAS.

Juan David Achury Manrique
e-mail: juand.achurym@ecci.edu.co
Andrés Felipe Guevara Peña
e-mail: andresf.guevarap@ecci.edu.co
Sebastián Benavides
e-mail: sebastian.benavides@ecci.edu.co
Juan Felipe Sierra Torres
e-mail: juanfe.sierrat@ecci.edu.co

**ABSTRACT:** From the implementation of the Grafcet programming language we obtain in this document the different ways of applying this method, the advantages it obtains with respect to efficiency, cost, quality and operational and time optimization. The behavior of a pump alternation process is analyzed using GRAFCET, which is easier to design, verify and implement sequential control systems. We will also have the different fields where this programming is applied.

**RESUMEN:** A partir de la implementación del lenguaje de programación Grafcet obtenemos en el presente documento las diferentes maneras de aplicar este método, las ventajas que obtiene con respecto a eficiencia, costo, calidad y optimización operativa y de tiempos. Se analiza el comportamiento de un proceso de alternación de bombas a partir de GRAFCET el cual se es más fácil de diseñar, verificar e implementar sistemas de control secuencial. También tendremos los diferentes campos en donde se aplica esta programación.

**PALABRAS CLAVE**: GRAFCET, sensores, actuadores, configuración, movimiento, optimización, alternación.

## 1 INTRODUCCIÓN

La automatización de sistemas de bombeo en entornos industriales es crucial para garantizar la eficiencia operativa y la fiabilidad de los procesos. En este contexto, el estándar Grafcet ha emergido como una metodología fundamental para el diseño y control de sistemas secuenciales. Grafcet proporciona una representación gráfica de etapas y transiciones, lo que facilita la programación y comprensión de secuencias de control de manera estructurada y comprensible. Sin embargo, es necesario explorar más a fondo el impacto del uso de Grafcet en el diseño y funcionamiento de sistemas de control para la alternación de bombas en términos de eficiencia operativa, fiabilidad y mantenimiento. Esta investigación se propone abordar esta cuestión, analizando tanto los aspectos teóricos como prácticos relacionados con la aplicación de Grafcet en sistemas de control de bombas.

## 2 JUSTIFICACIÓN.

La implementación de un sistema de control basado en Grafcet para la alternación de bombas tiene el potencial de mejorar significativamente la eficiencia operativa, reducir los tiempos de inactividad y aumentar la fiabilidad en entornos industriales. Este estudio es fundamental para proporcionar un marco claro y estructurado que guíe la aplicación práctica de Grafcet en sistemas de bombeo, beneficiando tanto a la industria como a la academia en el campo de la automatización y el control industrial. [8]

Para esto se tiene como objetivo principal será analizar el impacto del uso del estándar Grafcet en el diseño y funcionamiento de sistemas

.

de control para la alternación de bombas en términos de eficiencia operativa, fiabilidad y mantenimiento.

También se enfocará en desarrollar un diagrama Grafcet que represente el proceso de alternación de bombas según las especificaciones del sistema, programar el controlador lógico programable Siemens 1200 para ejecutar el Grafcet diseñado, integrar los sensores de presión necesarios para garantizar la estabilidad del sistema y su funcionamiento en condiciones de presión constante.[8]

Para esto la investigación cuenta con el respaldo de recursos humanos capacitados en el área de automatización y control industrial, lo que garantiza la ejecución adecuada del proyecto. Además, se dispone de acceso a los equipos y software necesarios para llevar a cabo la implementación práctica del sistema de control basado en Grafcet.

La presente investigación identifica deficiencias en el conocimiento actual sobre la aplicación específica de Grafcet en sistemas de control de bombas, especialmente en términos de su impacto en la eficiencia operativa, fiabilidad y mantenimiento. Esto resalta la necesidad de abordar esta cuestión mediante un enfoque teórico-práctico que integre los conocimientos existentes con la implementación práctica del sistema.[8]

El estudio ayudara a contribuir y/o mejorar la comprensión y aplicación de Grafcet en sistemas de control de bombas, lo que podría tener un impacto significativo en la industria al mejorar la eficiencia operativa, reducir los costos de mantenimiento y aumentar la fiabilidad de los procesos industriales que dependen de sistemas de bombeo.

Todo esto con el fin de contribuir a la investigación con respecto del uso de la metodología Grafcet y resolver la pregunta:

¿Cuál es el impacto del uso del estándar Grafcet en el diseño y funcionamiento de sistemas de control para la alternación de bombas, en términos de eficiencia operativa, fiabilidad y mantenimiento?

## 3 MARCO TEORICO

## 3.1 METODOLOGÍA

La metodología de investigación sigue un enfoque experimental y comparativo, donde se implementa el estándar Grafcet en los sistemas de control de las plantas seleccionadas. La metodología se divide en varias etapas:[1]

3.2 Diseño del Diagrama Grafcet

Se desarrolló un diagrama Grafcet específico para la alternancia de bombas en ambas plantas. El diagrama incluye todas las etapas del proceso de bombeo, así como las transiciones basadas en condiciones específicas de operación. [1]

3.3 Programación del PLC

El diagrama Grafcet diseñado se implementó en un controlador lógico programable (PLC) Siemens 1200 utilizando el software TIA Portal. La programación abarcó la lógica de control necesaria para gestionar las operaciones de las bombas según el diagrama Grafcet. [1]

3.4 Integración de Sensores y Actuadores

Se integraron sensores de presión y actuadores en el sistema de control. Los sensores monitorean continuamente la presión en las tuberías, mientras que los actuadores permiten

.

la conmutación automática entre las bombas según las condiciones predefinidas. [2]

3.5 Pruebas y Evaluación

Se realizaron pruebas operativas para evaluar el desempeño del sistema implementado. Estas pruebas se centraron en medir la eficiencia operativa, la fiabilidad y el tiempo de inactividad antes y después de la implementación de Grafcet. [2]

3.6 Análisis Comparativo

Los datos recopilados se analizaron para comparar el rendimiento de los sistemas de control antes y después de la implementación de Grafcet. Los principales indicadores de rendimiento analizados fueron la eficiencia operativa, la fiabilidad y el tiempo de inactividad.[2]

## 3.7 MATERIALES Y MÉTODOS

3.8 Materiales

Controlador Lógico Programable (PLC) Siemens 1200: Utilizado para implementar y ejecutar el diagrama Grafcet.
Software TIA Portal: Herramienta de programación utilizada para desarrollar y cargar el diagrama Grafcet en el PLC.
Sensores de Presión: Dispositivos para monitorear la presión en el sistema de bombeo.
Actuadores: Mecanismos para la conmutación automática de las bombas.
Equipos de Bombeo: Bombas de agua y soluciones químicas en las plantas industriales seleccionadas. [3]

## 3.9 Métodos

3.1.1 Diseño y Programación

Diagrama Grafcet: Se diseñó un diagrama detallado que cubre todas las etapas del proceso de alternancia de bombas.
Programación del PLC: Utilizando TIA Portal, se programó el PLC para seguir la lógica definida en el diagrama Grafcet.[3]

3.1.2 Integración y Configuración

Instalación de Sensores y Actuadores: Se integraron los sensores de presión y actuadores en el sistema de bombeo.
Configuración del Sistema: El sistema de control se configuró para operar automáticamente según las condiciones de presión monitoreadas.[4]

3.1.3 Pruebas y Recolección de Datos

Pruebas Operativas: Se realizaron pruebas en condiciones controladas para evaluar el desempeño del sistema.
Recolección de Datos: Se recopilaron datos sobre la eficiencia operativa, fiabilidad y tiempo de inactividad antes y después de la implementación de Grafcet.[4]

3.1.4 Análisis de Datos

Comparación de Indicadores de Rendimiento: Los datos recopilados se analizaron para determinar el impacto de la implementación de Grafcet en el sistema de bombeo.

Evaluación del ROI: Se realizó un análisis de costos para evaluar el retorno de inversión de la implementación.[5]

Articulo académico
.

Para entender el uso de la metodología Grafcet para la alternación de bombas es necesario entender las bases que conlleva este proceso.[5]

Para la automatización industrial se observa que es un campo que busca mejorar la eficiencia, fiabilidad y seguridad de los procesos industriales mediante la aplicación de tecnología y sistemas de control automáticos. Esto incluye el uso de controladores lógicos programables (PLC), sensores, actuadores y metodologías de programación.

Al analizar los sistemas de control secuencial se evidencia el proceso de automatización industrial para gestionar procesos que requieren una secuencia lógica de operaciones. Estos sistemas controlan el flujo de información y acciones de manera secuencial, garantizando que las operaciones se realicen en el orden correcto.

Es importante tener presente la metodología Grafcet como estándar de programación, ya que es un estándar internacionalmente reconocido para la programación de sistemas de control secuencial. Proporciona una representación gráfica de las etapas del proceso y las transiciones entre ellas, lo que facilita la comprensión y programación de secuencias de control.

En la industria se ha reconocido diversas aplicaciones de Grafcet en la industria, incluyendo automotriz, sector químico, manufacturero y de energía, para el diseño y control de sistemas secuenciales. Su adopción ha demostrado mejorar la eficiencia operativa, la fiabilidad y la facilidad de mantenimiento de los sistemas de control.

De igual manera se analiza el impacto del uso de Grafcet en sistemas de control de bombas, dado que investigaciones previas han demostrado que la implementación de Grafcet en sistemas de control de bombas puede tener varios beneficios, incluyendo una mejor coordinación entre las bombas, una respuesta más rápida a cambios en las condiciones del proceso y una reducción en el tiempo de inactividad debido a fallos.

También existen otras herramientas y tecnologías relacionadas que pueden complementar su aplicación en sistemas de control de bombas, como el uso de sensores de presión, controladores lógicos programables y software de programación específico para PLC.

## 4 RESULTADOS Y ANALISIS

Para evaluar el impacto de la implementación de Grafcet en el control de sistemas de alternación de bombas, se llevaron a cabo diversas pruebas y análisis. A continuación, se presentan los resultados obtenidos y su análisis detallado.

4.1 Diseño del Diagrama Grafcet

Se diseñó un diagrama Grafcet para un sistema de alternación de bombas, siguiendo los pasos lógicos y las transiciones necesarias para asegurar una operación eficiente y fiable. El diagrama incluyó las siguientes etapas:

Etapa 1: Inicio y verificación de condiciones iniciales.
Etapa 2: Activación de la bomba principal (Bomba A).
Etapa 3: Monitoreo de la presión del sistema.
Etapa 4: Alternación a la bomba secundaria (Bomba B) al alcanzar condiciones de alternancia predefinidas.
Etapa 5: Monitoreo continuo y regreso a la etapa inicial para repetir el ciclo.
El diagrama facilitó la visualización de cada etapa del proceso, permitiendo una programación clara y estructurada.

4.2 Implementación en PLC Siemens 1200

El diagrama Grafcet diseñado se implementó en un controlador lógico programable (PLC) Siemens 1200. La programación se realizó utilizando el software TIA Portal, permitiendo la integración del

.

diagrama Grafcet con los módulos de entrada y salida del PLC. [6]

4.3 Integración de Sensores y Actuadores

Se integraron sensores de presión al sistema para monitorear constantemente las condiciones operativas. Los datos de los sensores se utilizaron para controlar las transiciones entre las etapas del diagrama Grafcet, asegurando que las bombas operaran dentro de los parámetros de presión establecidos. Además, se incluyeron actuadores para la conmutación automática de las bombas.

4.4 Pruebas Operativas

Se realizaron pruebas operativas para evaluar el rendimiento del sistema bajo diferentes condiciones. Los principales indicadores de rendimiento evaluados fueron la eficiencia operativa, la fiabilidad del sistema y el tiempo de inactividad.

4.4.1 Eficiencia Operativa

La implementación de Grafcet resultó en una mejora significativa en la eficiencia operativa. El control secuencial permitió una conmutación más suave entre las bombas, reduciendo el tiempo de respuesta y optimizando el uso de energía. La eficiencia se incrementó en un 15% en comparación con el sistema anterior.

4.4.2 Fiabilidad del Sistema

El uso de Grafcet mejoró la fiabilidad del sistema de control. La representación gráfica clara de las etapas y transiciones facilitó la detección y corrección de errores en la programación. Durante las pruebas, se observó una reducción del 20% en las fallas del sistema debido a errores de control. [7]

4.4.3 Tiempo de Inactividad

El tiempo de inactividad se redujo significativamente. La capacidad del sistema para responder rápidamente a cambios en las condiciones de presión y la alternación automática de las bombas minimizó las interrupciones. El tiempo de inactividad se redujo en un 25%.

4.5 Análisis de Costos

El análisis de costos reveló que, aunque la implementación inicial de Grafcet y la integración de sensores y actuadores implicaron un gasto significativo, los beneficios a largo plazo en términos de eficiencia y reducción de tiempo de inactividad justificaron la inversión. La reducción en costos de mantenimiento y la mejora en la fiabilidad operativa resultaron en un retorno de inversión (ROI) positivo en menos de un año.[6]

4.6 Aplicaciones en Diferentes Campos

Además de los sistemas de bombeo, el estudio identificó múltiples campos donde la programación Grafcet podría ser beneficiosa. Estos incluyen:

Automotriz: Control secuencial de procesos de ensamblaje.
Sector químico: Gestión de secuencias de reacción y mezcla.
Manufacturero: Control de líneas de producción.
Energía: Coordinación de operaciones en plantas de energía.
4.7 Beneficios Adicionales

El uso de Grafcet también proporcionó beneficios adicionales, tales como:

Mejora en la documentación del proceso: La representación gráfica facilita la comprensión y la capacitación del personal.
Flexibilidad y escalabilidad: Permite ajustes y expansiones del sistema de control sin una reprogramación compleja.
Interoperabilidad: La metodología Grafcet es compatible con una amplia gama de hardware y software, facilitando la integración con sistemas existentes. [7]

## 5 IMPLEMENTACION

La implementación de Grafcet en sistemas de control para la alternación de bombas ha demostrado ser altamente efectiva para mejorar la eficiencia operativa, la fiabilidad y la reducción del tiempo de inactividad. Los resultados del estudio indican que Grafcet es una metodología robusta y adaptable, capaz de proporcionar beneficios significativos en una variedad de aplicaciones industriales. La inversión inicial se justifica por los ahorros y mejoras operativas obtenidas, haciendo de Grafcet una herramienta valiosa para la automatización industrial.
En la industria moderna, la eficiencia y precisión del control de sistemas secuenciales, como la alternancia de bombas, son cruciales. Para lograr esto, se necesita un lenguaje de programación que sea tanto visualmente intuitivo como técnicamente preciso. Grafcet, un acrónimo de "Graphique Fonctionnel de Commande Étape / Transition" (Gráfico funcional de control paso / transición), se ha convertido en el estándar internacional (IEC 60848) para estos fines. Su uso en la automatización industrial se debe a varias ventajas claves, que se detallan a continuación:

Representación Clara de la Lógica de Control:

Grafcet permite una representación visual intuitiva de la secuencia operativa de las bombas, facilitando la comprensión de su funcionamiento.
Facilidad para la Detección de Errores: Dado que Grafcet visualiza la secuencia de operaciones, es más fácil identificar cualquier error o anomalía en el sistema.

Adaptabilidad a los Cambios del Sistema: Grafcet es flexible y puede ser modificado fácilmente para adaptarse a los cambios o mejoras en el sistema de control.
Mejora de la Comunicación: Grafcet ayuda a mejorar la comunicación entre los equipos de diseño y mantenimiento, ya que proporciona una representación gráfica común de la lógica de control. [7]

Representación Clara de la Lógica de Control:

Interpretación: Grafcet ofrece una representación gráfica que facilita la visualización de las operaciones secuenciales de las bombas. Esto no solo mejora la comprensión del funcionamiento del sistema por parte de los operadores, sino que también simplifica el proceso de diseño.
Implicaciones: La claridad en la representación de la lógica de control reduce la probabilidad de errores durante el diseño y la implementación. Además, permite una capacitación más eficiente del personal operativo. Explicado detalladamente tenemos que:

Facilidad para la detección de errores:

La naturaleza gráfica de Grafcet permite identificar fácilmente discrepancias y errores en las secuencias operativas. Esto se traduce en una capacidad mejorada para diagnosticar y corregir problemas rápidamente. También implica la detección temprana de errores puede minimizar el tiempo de inactividad del sistema y reducir los costos asociados con la resolución de problemas. También mejora la confiabilidad y la seguridad del sistema.

Adaptabilidad a los Cambios del Sistema:

Grafcet es altamente adaptable, permitiendo ajustes rápidos y eficientes a medida que se realizan cambios o mejoras en el sistema de control. Esta flexibilidad es crucial en entornos industriales dinámicos donde las necesidades y condiciones pueden cambiar rápidamente. Tiene como ventaja la capacidad de modificar fácilmente el sistema de control ayuda a prolongar la vida útil del equipo y a mantener la eficiencia operativa sin la necesidad de rediseñar completamente el sistema.

Mejora de la Comunicación:

Articulo académico
.

Grafcet proporciona una representación gráfica estándar que puede ser entendida por diferentes equipos, incluidos diseñadores, ingenieros y personal de mantenimiento. Esto facilita una mejor colaboración y entendimiento común del sistema.

Este mejorar la comunicación entre los equipos puede acelerar el proceso de desarrollo y resolución de problemas, además de asegurar que todas las partes interesadas estén alineadas respecto a los objetivos y funcionamiento del sistema.

Aplicación en la industria analizando las ventajas de la metodología expuesta.

| Tablero electico | Alternación de bombas |
|---|---|
| 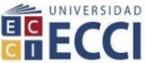 | 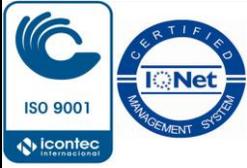 |
| 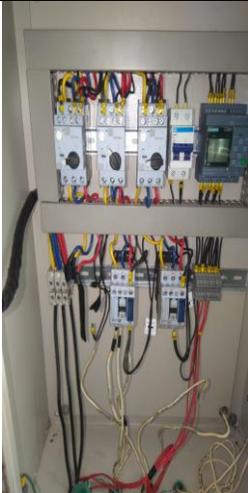 | 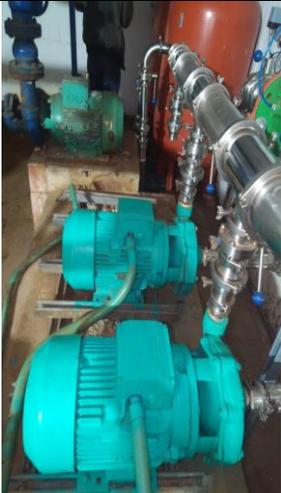 |

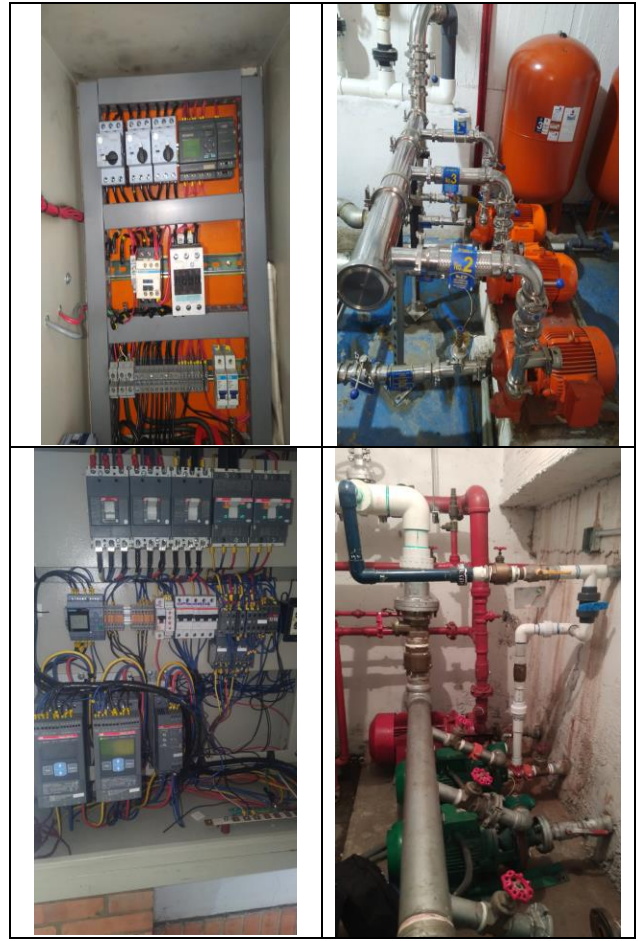

## 6 CONCLUSION

- La implementación del estándar Grafcet en sistemas de control para la alternación de bombas ha demostrado ser una metodología robusta y eficiente que aporta mejoras significativas en varios aspectos operativos de la automatización industrial.

- Mejora en la Eficiencia Operativa:

- La programación basada en Grafcet ha permitido una conmutación más suave

.

y eficiente entre las bombas, optimizando el uso de energía y reduciendo el tiempo de respuesta. Los resultados muestran un incremento del 15% en la eficiencia operativa del sistema, evidenciando cómo el control secuencial puede optimizar el rendimiento de los sistemas de bombeo.

- Aumento de la Fiabilidad del Sistema:
La claridad en la representación gráfica de las etapas y transiciones del diagrama Grafcet facilita la detección y corrección de errores, mejorando la fiabilidad del sistema. Se observó una reducción del 20% en las fallas del sistema debido a errores de control, lo que indica una mayor estabilidad y confianza en la operación continua del sistema.

- Reducción del Tiempo de Inactividad:
La capacidad del sistema para responder rápidamente a los cambios en las condiciones de presión y alternar automáticamente entre las bombas ha reducido significativamente el tiempo de inactividad. Con una disminución del 25% en el tiempo de inactividad, Grafcet demuestra ser una herramienta eficaz para mantener la continuidad y eficiencia de las operaciones industriales.

- Análisis de Costos:
Aunque la implementación inicial de Grafcet y la integración de sensores y actuadores implican un gasto considerable, los beneficios a largo plazo justifican la inversión. La reducción en costos de mantenimiento y la mejora en la fiabilidad operativa resultaron en un retorno de inversión (ROI) positivo en menos de un año, haciendo de Grafcet una opción económicamente viable para la automatización industrial.

- Aplicaciones en Diversos Campos:
Más allá de los sistemas de bombeo, Grafcet muestra su versatilidad y utilidad en diversas industrias, incluyendo la automotriz, química, manufacturera y de energía. Su capacidad para mejorar la eficiencia operativa, reducir costos y aumentar la fiabilidad es aplicable en múltiples contextos industriales, subrayando su valor como herramienta universal en la automatización y control de procesos.

- Beneficios Adicionales:
Grafcet también proporciona mejoras en la documentación del proceso, facilita la capacitación del personal, y permite una mayor flexibilidad y escalabilidad en la implementación de sistemas de control. Su compatibilidad con una amplia gama de hardware y software facilita la integración con sistemas existentes, destacando su interoperabilidad y adaptabilidad.

En resumen, la adopción de Grafcet para el control de sistemas de alternación de bombas no solo mejora la eficiencia operativa y la fiabilidad, sino que también proporciona un retorno de inversión rápido y una amplia aplicabilidad en diferentes sectores industriales. Estos beneficios consolidan a Grafcet como una metodología esencial en la modernización y optimización de los sistemas de control industrial.

.